\documentclass[12pt,a4paper]{article}
\usepackage{a4wide}
\usepackage[centertags]{amsmath}
\usepackage{amssymb}
\usepackage{epsfig}
\usepackage{cite}
\usepackage{ulem}
\usepackage{float}
\def\be{\begin{equation}}
\def\ee{\end{equation}}
\def\bea{\begin{eqnarray}}
\def\eea{\end{eqnarray}}

\begin{document}

\thispagestyle{empty}

\begin{center}
{\Large \bf Nontrival Cosmological Constant in Brane Worlds with Unorthodox  Lagrangians } 
\end{center}

\vspace*{1cm}

\centerline{\large Stefan F\"orste, Hans Peter Nilles, Ivonne Zavala }

\vspace{1cm}

\begin{center}{\it
Bethe Center for Theoretical Physics\\
{\footnotesize and}\\
Physikalisches Institut der Universit\"at Bonn,\\
Nussallee 12, 53115 Bonn, Germany}
\end{center}

\vspace*{1cm}

\centerline{\bf Abstract}
\vskip .3cm
In self-tuning brane-world models with extra dimensions, large
contributions to the cosmological constant are absorbed into the
curvature of extra dimensions and consistent with flat 4d geometry. 
In models with conventional Lagrangians fine-tuning is needed
nevertheless to ensure a finite effective Planck mass. Here, we
consider a class of models with non conventional Lagrangian in which 
known problems can be avoided. Unfortunately these models are found to
suffer from tachyonic instabilities. 
An attempt to cure these instabilities leads to the prediction
of a positive cosmological constant, which in turn 
needs a fine-tuning to be consistent with observations.

\vskip .3cm

\newpage

\section{Introduction}

Quantum corrections to the cosmological constant are of the order
$M_p^4$ where $M_p$ denotes the Planck mass. To achieve agreement with
observations that the cosmological constant is less than $10^{-120}
M_p^4$ requires a severely fine-tuned bare value. (With supersymmetry
broken below the Planck scale the fine-tuning is slightly relaxed but
cannot be sufficiently
removed.) Reviews on the cosmological constant problem are  e.g.\ listed in
\cite{Weinberg:1988cp,Witten:2000zk,Binetruy:2000mh,Ellwanger:2002cd}. 

About ten years
ago, the fine-tuning problem was revisited in the context of
brane-worlds. The basic idea is that some extra directions can be
probed only by gravitational interactions and are thus not visible to
us. Vacuum energy created by matter and the other interactions could
curve the extra dimension and would not be observable. Famous
examples in which such a picture is realised are the Randall-Sundrum (RS) models \cite{Randall:1999ee,Randall:1999vf}. There, however, the
fine-tuning appears in matching conditions between bulk and brane 
parameters \cite{Randall:1999ee,Randall:1999vf,DeWolfe:1999cp}. 
Later it was proposed to 
add a bulk scalar field in order to avoid this fine-tuning
\cite{Kachru:2000hf,ArkaniHamed:2000eg}. Indeed, the matching
condition between bulk and brane parameters just fixes integration
constants and thus self-tuning is apparently achieved. However, 
to avoid singularities and to get a finite
value for the Planck mass one needs to cut-off the extra
dimension. This can be done consistently only by adding branes with
fine-tuned tensions \cite{Forste:2000ps,Forste:2000ft}. It can be
shown that this problem persists for quite general bulk potentials of
the scalar \cite{Csaki:2000wz}.

The authors of \cite{Kim:2001ez} proposed an interesting model which 
circumvents
all these known problems. It has an unconventional kinetic term,
expansion around a vacuum configuration results in a higher derivative
model. (For subsequent work on this model see \cite{Kim:2010fb} and
references therein.)
A quite different analysis also hints at unconventional bulk
Lagrangians to be needed for a working self-tuning mechanism
\cite{Antoniadis:2010sz,Antoniadis:2010ik}. (Their conditions however seem to
differ from the ones in \cite{Kim:2001ez} and the ones considered in
the present article.)  

We will describe a class of models in which a bulk scalar has an
unconventional kinetic term. These models can also be seen as a natural generalisation of RS \cite{Randall:1999ee,Randall:1999vf} and \cite{Kachru:2000hf,ArkaniHamed:2000eg}, which are contained in our set up. These models have a single brane,
matching conditions just fix integration constants, singularities are avoided 
and the Planck mass
will be finite without introducing extra branes. 
A stability analysis
shows that perturbations depending on the extra coordinate are indeed damped in one class of models, while to achieve 
damping of fluctuations depending on the
visible directions fine-tuning has to be reintroduced. 
The class of
models discussed here contains the model in \cite{Kim:2001ez} in its
dual formulation \cite{Choi:2002pp}. 
In this case, perturbations depending on the visible coordinates are
damped, while  fluctuations depending on the 
extra dimension give rise to tachyonic Kaluza Klein (KK)
masses\footnote{A related argument for the  
instability of that particular case was presented in
\cite{Medved:2001ad}.}. 
Attempts to cancel such negative mass-squareds lead to the 
prediction that the cosmological constant has to be nonzero and
positive. Unfortunately the size of the cosmological constant
is again a subject of fine-tuning.

We will not consider more than one extra dimension. Examples with two
extra dimensions are discussed in
\cite{Chen:2000at,Carroll:2003db,Navarro:2003vw,Aghababaie:2003wz,Aghababaie:2003ar,Nilles:2003km,Lee:2003wg,Garriga:2004tq,Burgess:2004dh,Wetterich:2010kd}
(where especially \cite{Nilles:2003km,Lee:2003wg,Garriga:2004tq}
provide a critical evaluation of self-tuning proposals).

The paper is organised as follows. In the next section we construct
self-tuning solutions with one extra dimension. The kinetic term of a
bulk scalar is raised to a real power $\alpha$. Conditions on $\alpha$
from imposing a finite effective Planck mass are derived in section
3. As a crosscheck we compute the effective cosmological constant in
section 4 and give an estimated lower bound on the size of the
extra dimension. In section 5 we show that imposing stability
enforces non vanishing effective cosmological constant and 
reintroduces a fine-tuning condition. Section 6 discusses nearby
curved solution. In section 7, we argue that the problems persist
upon adding a bulk cosmological constant. In section 8 we relate
our class of models to models with a three-form potential via electric
magnetic  duality. In section 9 we summarize our results.

\section{``Self-Tuning'' with Unconventional Lagrangian}

The system we consider consists of a bulk action and a brane action
$$
S = S_{\text{bulk}}+S_{\text{brane}},
$$
with
\begin{equation}\label{eq:bulkaction}
S_{\text{bulk}}=\int d^4x dy \sqrt{-G}\left(
  \frac{R}{2\kappa_5^2} - \lambda \left( \partial_M \Phi\partial^M
    \Phi^\dagger\right)^\alpha \right)
\end{equation}
and
\begin{equation}
S_{\text{brane}} = -\int d^4 x\sqrt{-g}f\left(
  \Phi\right)_{\left| y=0\right.} .
\end{equation}
Here, $G_{MN}$ is the five dimensional metric. The fifth coordinate is
called $y$ and the 4d coordinates are $x^\mu$, $\mu =0,\ldots ,3$. 
Moreover, the parameters $\lambda$ and $\alpha$ are for the moment arbitrary. 
Notice that the self-tuning system discussed in \cite{Kachru:2000hf}  can be recovered from (\ref{eq:bulkaction}) by taking $\lambda = 4/3$ and $\alpha=1$, which indeed corresponds to a bulk scalar field with standard kinetic term. Moreover, the RS solution \cite{Randall:1999ee,Randall:1999vf} is recovered for $\alpha =0 $ and $\lambda =1$, which corresponds simply to a bulk cosmological constant. Thus it is reasonable to consider as an interpolating scheme
the more general system above with arbitrary $\alpha$.

The brane is taken to 
be localised within the fifth dimension at $y=0$. Then the induced
metric on the brane reads
\begin{equation}
g_{\mu\nu} = G_{MN}\delta^M _\mu \delta^N _{\nu\left| y=0\right.} , \quad \mu ,\nu =
  0,\ldots ,3.
\end{equation}
We have taken a complex scalar to have single valuedness of the action
under $\Phi \to e^{2\pi \mbox{\scriptsize i}} \Phi$. The differences to
  a real field turn out to be less important. For the metric
  we take an ansatz compatible with 4d Minkowski isometry
\begin{equation}\label{eq:mansatz}
ds^2 = a\left(y\right)^2\eta_{\mu\nu}\,dx^\mu dx^\nu + dy^2 ,
\end{equation}
where $\eta_{\mu\nu} = diag\left( -1,1,1,1\right)$ is the Minkowski
metric. The equations of motion are
\begin{equation}\label{eq:einst55}
6 \left( \frac{a^\prime}{a}\right) ^2 = \kappa_5^2 \lambda \left( 2\alpha
  -1\right) \left| \Phi^\prime\right|^{2\alpha}
\end{equation}
from the $yy$ component of Einstein's equation. We allow at most for
$y$ dependence and denote the corresponding derivatives with a
prime. The $\mu\nu$ components of Einstein's equation give rise to
\begin{equation}
3 \frac{a^{\prime\prime}}{a} = -\frac{\lambda \kappa_5^2}{2}\left( 2
  \alpha +1\right)\left| \Phi^\prime\right|^{2\alpha} - f\left(
  \Phi\right)\delta\left( y\right) \kappa_5^2 .
\end{equation}
Finally, the equation for the complex scalar reads
\begin{equation}\label{eq:scalar}
\frac{\lambda\alpha}{a^4}\left( a^4 \left| \Phi^\prime\right|^{2\left(
      \alpha -1\right)} \Phi^\prime\right)^\prime = \frac{\partial
  f}{\partial \Phi^\dagger}\delta\left( y\right) .
\end{equation}
We proceed now to solve these equations. From the Einstein equation
(\ref{eq:einst55}) we get
\begin{equation}
\frac{a^\prime}{a} = \pm A \left| \Phi^\prime\right|^\alpha
\end{equation}
with 
\begin{equation}
A = \sqrt{\frac{\kappa_5^2 \lambda \left( 2 \alpha -1\right)}{6}} .
\end{equation}
Reality of $a$ forces us to take the parameters $\lambda$ and
$\alpha$ such that the argument under the square root is real and
positive, i.e.\
\begin{equation}\label{eq:areal}
\lambda\left( 2\alpha -1\right) > 0
\end{equation}
Therefore we have two types of solutions: 
a) $\alpha>1/2, \,\lambda>0$ (the solutions in \cite{Kachru:2000hf} fall into this class and one could then suspect the appearance of singularities) and  
b) $\alpha< 1/2,\, \lambda<0$. As we see below case b) is the most interesting one from the point of view of a realisation of the self-tuning mechanism. 

We pick the branch of the root such that $A>0$. Using this
in the scalar equation (\ref{eq:scalar}) and combining with its
complex conjugate we obtain
\begin{equation}
\lambda \alpha \left| \Phi^\prime\right|^{2\alpha}\left(
  \frac{\Phi^{\prime\prime}}{\Phi^\prime}
    -\frac{{\Phi^\dagger}^{\prime\prime}}{{\Phi^\dagger}^\prime}\right)
  = \left( {\Phi^\dagger}^\prime\frac{\partial f}{\partial
      \Phi^\dagger}-\Phi^\prime \frac{\partial f}{\partial
      \Phi}\right)\delta\left( y\right).
\end{equation}
For $y\not= 0$ this is solved by 
\begin{equation}
\Phi^\prime = e^{\mbox{\scriptsize i}\rho}\varphi^\prime ,
\end{equation}
where $\rho$ is a real integration constant (which can jump at $y=0$) and
$\varphi$ is a real $y$ dependent field. Further we disregard the
possibility of having constant $\Phi$ since that leads to a constant
warp factor. A constant warp factor either results in a divergent
effective Planck mass or one has to introduce more branes to
compactify the fifth direction by cutting it off. The equation for the
remaining real scalar $\varphi$ is
\begin{equation}
\lambda \alpha \left| \varphi\right|^{2\alpha}\left( \left( 2 \alpha
    -1\right) \frac{\varphi ^{\prime\prime}}{\varphi^\prime} \pm
  4A\left| \varphi^\prime\right|^\alpha\right) = e^{-\mbox{\scriptsize
    i}\rho} \varphi^\prime \frac{\partial f}{\partial \Phi^\dagger}
\delta\left( y\right) .
\end{equation}
For $y\not= 0$ this leads to
\begin{equation}\label{eq:scaloneint}
e^{-\alpha \log \left| \varphi^\prime\right|}= \left|
  \varphi^\prime\right|^{-\alpha} = \pm \frac{4\alpha A\, y}{2\alpha -1}
+ C_1,
\end{equation}
where $C_1$ is a real integration constant which can again change at $y=0$.
The left hand side of (\ref{eq:scaloneint}) is never
negative. Curvature singularities at finite $y$ can be avoided if the
right hand side does not change sign. So, we take 
\begin{equation}
C_1 \geq 0. 
\end{equation}
The first term on the right hand side of (\ref{eq:scaloneint}) should
neither be negative.  So, for 
instance if (this will turn out to be the case of interest below)
\begin{equation} 
\frac{\alpha}{2\alpha -1}<0 
\end{equation}
we take the lower sign for $y>0$ and the upper sign for $y<0$. The
complete solution for the complex scalar is then
\begin{equation}
\Phi = -\frac{\Theta\left( y\right) \left( 2\alpha
    -1\right)}{4 A\left(\alpha -1\right)}e^{\mbox{\scriptsize i}
  \rho}\left( -\frac{4\,\alpha A}{2\alpha 
    -1}\left| y\right| +C_1\right)^{\frac{\alpha -1}{\alpha}} +C_2 ,
\end{equation}
where $C_2$ is a complex integration constant and $\Theta\left(
  y\right)$ denotes the Heaviside step function. Solving Einstein's
equations for $y\not=0$ yields the warp factor
\begin{equation}\label{eq:warpsol}
a =\left( C_3\left( - \frac{4\,\alpha A}{2\alpha -1}\left| y\right|
    +C_1\right)\right)^{\frac{2\alpha -1}{4\alpha}} .
\end{equation}
Analysing our set of equations at $y=0$ one finds that $\Phi$ and $a$
should be continuous. Denoting the value of an integration constant at
$y\gtrless 0$ by a superscript $\pm$ we find from imposing
continuity of $\Phi$ and $a$
\begin{align}
-\frac{2\alpha -1}{4A\left(\alpha -1\right)}e^{\text{i} \rho^{+}}\left(
    C_1^{+}\right)^{\frac{\alpha -1}{\alpha}} + C_2^{+} & =
\frac{2\alpha -1}{4A\left(\alpha -1\right)}e^{\text{i} \rho^{-}}\left(
    C_1^{-}\right)^{\frac{\alpha -1}{\alpha}} + C_2^{-} , \\
\left( C_3^+ C_1^+\right)^{\frac{2\alpha -1}{4\alpha}} &=
\left( C_3^- C_1^-\right)^{\frac{2\alpha -1}{4\alpha}} .\label{eq:contc1c3}
\end{align}
First derivatives have to jump by a finite amount such that second
derivatives reproduce the delta function on the right hand side of the
equations of motion. This leads to the following jump conditions on
the integration constants
\begin{align}
\lambda \,\alpha \left( e^{\text{i} \rho^+}\left(
    C_1^+\right)^{\frac{1-2\alpha}{\alpha}} -
e^{\text{i} \rho^-}\left(
    C_1^-\right)^{\frac{1-2\alpha}{\alpha}}\right) &= \frac{\partial
  f}{\partial \Phi^\dagger} , \\
\frac{1}{C_1^+} +\frac{1}{C_1^-} = \frac{\kappa_5^2 f\left(
    \Phi\right)}{3 A}\label{eq:ccpos} .
\end{align}
From our earlier finding that $C_1 \geq 0$ and (\ref{eq:ccpos})  we see
that the solution with 4d Minkowski isometry exists for any value of a
positive cosmological constant on the brane. {\it This is the 
so-called ``self-tuning''
mechanism}. (The quotation marks indicate that the solution still has
to pass some consistency condition to be discussed subsequently.)

\section{Effective Planck Mass}

The effective four dimensional Planck mass, $M_p$, can be obtained as
follows. We
replace the 4d Minkowski metric in (\ref{eq:mansatz}) by a general
$x$ dependent metric, plug that into the bulk action
(\ref{eq:bulkaction}) and integrate over the fifth dimension. The
result contains a 4d Einstein-Hilbert term, the factor in front of
which yields the effective 4d gravitational coupling, $\kappa$. One obtains
\begin{equation}\label{Mplanck4}
\frac{1}{\kappa} = \frac{M_p^2}{8\pi} = \frac{1}{\kappa_5^2}\int dy\, a^2
\end{equation}
With our solution (\ref{eq:warpsol}) we get
\begin{align}
\frac{\kappa_5 ^2}{\kappa} = \int_{-y_c}^{y_c} dy \left( C_3\left( -
    \frac{\alpha 4 A}{2\alpha -1}\left| y\right| + C_1\right)\right)
^{\frac{2\alpha -1}{2\alpha}}& =\nonumber\\ &\hspace*{-1in}
\left. -\frac{2\alpha -1}{2A\left( 
      4\alpha -1\right)}\frac{\Theta\left( y\right)}{C_3}\left(
    C_3\left(  -
    \frac{\alpha 4 A}{2\alpha -1}\left| y\right| +
    C_1\right)\right)^{\frac{4\alpha
    -1}{2\alpha}}\right|^{y_c}_{y=-y_c} ,
\end{align}
where we have cut off the integration at $y=\pm y_c$. To
consistently do so one would need to introduce additional branes with
fine-tuned tension at those positions
\cite{Forste:2000ps,Forste:2000ft}. Such fine-tuning 
can be avoided while obtaining finite Planck mass when the limit
$y_c \to \infty$ is finite. This leads to the condition 
\begin{equation}\label{eq:arange}
0< \alpha <\frac{1}{4}.
\end{equation}
Thus we see that solution b) $\alpha< 1/2,\, \lambda<0$  is the only
one which could give rise to a successful self-tuning
mechanism\footnote{Note that in the present set up, an apparent wrong
  sign for the scalar field kinetic term, does not imply instabilities
  or the presence of ghosts. What needs to have the correct kinetic
  term are the fluctuations around the background solution, as we see
  below.}.  
Notice that the region between $1/4< \alpha < 1/2$ is excluded due to the requirement of finiteness of the Planck mass (see Fig.~1). We see below that adding a bulk cosmological constant can help in allowing this region. 
 
If  (\ref{eq:arange})  holds, the four dimensional gravitational coupling is given by
\begin{equation}\label{eq:4dmp}
\frac{\kappa_5^2}{\kappa} = \frac{2\alpha -1}{2A\left( 4 \alpha -1\right)}
\left( C_3 C_1\right)^{\frac{4\alpha -1}{2\alpha}}\left(
  \frac{1}{C_3^+} +\frac{1}{C_3^-}\right) ,
\end{equation}
where we used the continuity of $C_1C_3$ (\ref{eq:contc1c3}).

\section{Cross Check: Effective 4d Cosmological Constant}

Since 4d slices in our five dimensional brane world are flat, the
effective cosmological constant should vanish. It is computed as the
Lagrangian evaluated at the solution and integrated over the fifth
direction,
\begin{equation}
\Lambda_{\text{eff}} = \int_{-y_c}^{y_c} dy\, a^4\left( -
  \frac{4 a^{\prime\prime}}{\kappa_5^2 a} -\frac{6}{\kappa_5^2}\left(
    a^{\prime}{a}\right)^2 -\lambda\left| \Phi^\prime\right|^{2\alpha}\right)
-a^4 f_{\left| y=0\right.} .
\end{equation}
Plugging in our solution, taking carefully into account delta function
contributions to second derivatives, we find
\begin{equation}
\Lambda_{\text{eff}} = \left.\left( C_3 \right)^{\frac{2\alpha
    -1}{\alpha}}\frac{\lambda}{6A} \Theta\left( y\right) \left(
  -\frac{\alpha 4 A}{2\alpha -1}\left| y\right| +
  C_1\right)^{\frac{\alpha -1}{\alpha}}
\right|_{y=-y_c}^{y_c} + \frac{1}{3}\left( C_3
  C_1\right)^{\frac{2\alpha -1}{\alpha}} f .
\end{equation}
From the matching conditions we see that the
contribution at $y= \pm 0$ cancels and we are left with
\begin{equation}\label{eq:lameff}
\Lambda_{\text{eff}} \sim\left( y_c M_p\right)^{\frac{\alpha
    -1}{\alpha}} M_p^4 ,
\end{equation}
where we deduced the dependence on $M_p$ on dimensional grounds (with
a brane tension of order one in Planck units 
integration constants also contribute with that
order\footnote{Actually, $\lambda$ can be also dimensionful depending
  on the dimension assigned to $\Phi$. Therefore, one should re-scale
  $C_1$ and $C_3$ such that the $\lambda$ dependence drops out of
  (\ref{eq:ccpos}) and (\ref{eq:4dmp}). Since $A \sim \sqrt{\left|
      \lambda\right|}$ this amounts to $C_1 \to
  \sqrt{\left|\lambda\right|}C_1$ and $C_3 \to
  C_3/\sqrt{\left|\lambda\right|}$. With that one can see that
  $\Lambda_{\text{eff}}$ does not depend on  
  $\lambda$.\label{fn:lam}}). 
We see that with our choice (\ref{eq:arange}) we can take the limit
$y_c \to \infty$ and the effective cosmological constant
vanishes.

Since the observed cosmological constant is not exactly zero we could
put a lower bound on $y_c$. There are, however, some
subtleties. Just choosing $y_c$ such that the effective
cosmological constant computed here agrees with the observed value
would not be fully consistent since our 4d slices are still
flat. Further, once cutting off the space at some large value $\left|
  y\right| = y_c$ it is natural to add branes there whose not
fine-tuned tension would contribute as
$$ a^4 f\left( \Phi\right)_{\left| y = \pm y_c\right.}. $$
In section \ref{sec:curved} we will discuss consistent solutions with
a finite cosmological constant. For the moment let us assume that
cutting off the $y$ integration at $y_c$ without adding
fine-tuned brane contributions results in a consistent solution with an
effective cosmological constant given by (\ref{eq:lameff}). So, we
estimate a bound on $y_c$ as
\begin{equation}\label{eq:lamest}
y_c M_p> \left( 10^{120}\right)^{\frac{\alpha}{1-\alpha}} .
\end{equation}
Note that taking the limit $y_c\to \infty $ satisfies indeed the
condition above, thus giving a vanishing effective 4d cosmological
constant. We will come back to this point in Section \ref{sec:curved}
where we discuss nearby curved solutions.  

\section{Stability and Hidden Fine-Tuning}

Because of the unusual form of the Lagrangian one should ensure that
our self-tuning solution is stable against fluctuations in the scalar
field. We will do so in two steps. First, we consider fluctuations
which depend only on the fifth direction, $y$. It turns out that
these fluctuations are suppressed. For that reason, we take in a
second step fluctuations which depend only on directions $x^\mu$ of
4d space-time.
There, we have to add kinetic
terms to the brane to stabilise them. Canonically normalising the
kinetic terms on the brane we derive conditions on the coupling
$\lambda$. If we think of $\lambda$ as a superposition of a bare
quantity and quantum corrections both of them being of order one in
Planck units these conditions can be met only with extreme fine-tuning. 

So, first we take
\begin{equation}
\Phi = \Phi_{cl} + \delta \Phi\left( y\right)
\end{equation}
and expand our bulk Lagrangian till second order in the
fluctuation. The result for the second order term is
\begin{equation} 
- \lambda\frac{\alpha}{2} \left| \Phi_{cl}^\prime\right|^{2\left( \alpha
    -2\right)}\left( \delta {\Phi^\dagger}^\prime, \delta \Phi^\prime\right) M
\left( \begin{array}{c} \delta \Phi^\prime\\ \delta
    {\Phi^\dagger}^\prime\end{array}\right) , 
\end{equation}
where the two-by-two matrix $M$  is 
\begin{equation}
M =\left( \begin{array}{cc}
\alpha \left| \Phi_{cl}^\prime\right|^2 & 
\left( \alpha -1\right)
\left(\Phi_{cl}^\prime\right)^2 \\
\left( \alpha -1\right)
\left({\Phi_{cl}^\dagger}^\prime\right)^2 &
\alpha \left|\Phi_{cl}^\prime\right|^2\end{array}\right) ,
\end{equation}
with determinant
\begin{equation}
\det M = \left( 2\alpha -1\right) \left| \Phi_{cl}^\prime\right|^4 .
\end{equation}
From our condition ({\ref{eq:areal}) we see that $y$ dependence in
  scalar fluctuations is under control. 
For this reason we focus
  in the following on fluctuations depending only on non-compact
  coordinates, $x^\mu$, i.e.\
\begin{equation}
\Phi = \Phi_{cl} + \delta\Phi\left( x\right)
\end{equation}
An effective 4d Lagrangian will contain the following term quadratic
in the fluctuations
\begin{equation}\label{flucs}
{\cal L}_{4d} \supset \int^{y_c}_{-y_c} dy \left(
  -\lambda\right)\alpha a^2 \left| 
  \Phi^\prime_{cl}\right|^{2\alpha -2} \partial_\mu  \delta \Phi\left(
  x\right)^\dagger \partial^\mu \delta \Phi\left( x\right) . 
\end{equation}
Our condition (\ref{eq:areal}) and (\ref{eq:arange}) show that $x$
dependent fluctuations are unstable. By performing the $y$ integral
we find even a factor
\begin{equation}
{\cal L}_{4d} \supset \sim \left|
  \lambda\right|^{\frac{1}{\alpha}}\left( y_c
    M_p\right)^{\frac{3}{2\alpha}} M_p^{\frac{-5 +2\alpha\left( 1 +
        d_\Phi\right)}{\alpha} +2 - 2 d_{\Phi}} \partial_\mu\delta \Phi\left(
  x\right)^\dagger \partial^\mu \delta \Phi\left( x\right) 
\end{equation}
which diverges in the limit $y_c \to \infty$. Keeping $y_c$ finite thus gives rise to  $\Lambda_{eff}\ne 0$ unless we can stabilise the solution in some way (see below). Here, $d_\Phi$ denotes
the mass dimension of the scalar field\footnote{The mass dimension of
  $\Phi$ is fixed in terms of $\lambda$'s mass dimension, $d_\lambda$,
  and the parameter $\alpha$ according to $2\alpha\left( 1+
    d_\Phi\right) + d_\lambda = 5$.}. The $\lambda$ dependence can
be either established along the lines of footnote \ref{fn:lam} or by
using the symmetry under simultaneous re-scalings of $\left|\partial
  \Phi\right|^2$ and $\lambda$ of our original Lagrangian. The
dependence on the Planck mass follows from counting mass dimensions.

 Stabilisation of the solution can be done by adding a term
to the brane Lagrangian
\begin{equation}
{\cal L}_{\text{brane}} \to {\cal L}_{\text{brane}} - M_p^{2 -
  2d_\Phi}\partial_\mu  \Phi\left( 
  x\right)^\dagger \partial^\mu \Phi\left( x\right)
\end{equation}
where the factor is taken to be one in Planck units to avoid fine-tuning.    
This term has to cancel at least the unstable contribution from the
bulk leading to the condition
\begin{equation}
\tilde{\lambda}^{\frac{1}{\alpha}}\left(y_c M_p\right)^{\frac{3}{2\alpha}} <1 ,
\end{equation}
where 
$$ \tilde{\lambda} = \left|\lambda\right| M_p^{-5 + 2\alpha\left( 1 +
    d_\Phi\right)} $$
is a dimensionless quantity. Using our estimate on a lower bound for
$y_c$ (\ref{eq:lamest}) we find the condition
\begin{equation}\label{eq:hidft}
\tilde{\lambda}^{\frac{1}{\alpha}} < 10^{- \frac{180}{\left(
      1 -\alpha\right)}} .
\end{equation}
Note, that $\left|\lambda\right|^{1/\alpha}$ is the 
quantity which appears naturally when we balance bulk and brane
contributions to achieve stability.
Of course we could consider some small power of that 
quantity and improve the fine-tuning condition (\ref{eq:hidft}),  but
that would be as good as considering some small power of the ratio of
observed to computed cosmological constant.
So, for our range (\ref{eq:arange}) the fine-tuning turns out to be
even worse than the original fine-tuning of a bare cosmological
constant to cancel quantum corrections. But this is only a rough
estimate. We do not know how quantum corrections to our original
action look like and just assumed that they are of order one in Planck
units (such that no fine-tuning corresponds to $\tilde{\lambda} \sim
1$). Still it is clear that the condition on the coupling is related to
the small size of the observed value of the cosmological constant and
implies hidden fine-tuning. Using the relation (\ref{eq:lameff}) we
can also express the condition on $\lambda$ as
\begin{equation}
\tilde{\lambda}^{\frac{1}{\alpha}} < \left(\frac{
  \Lambda_{\text{eff}}}{M_p^4}\right) ^{\frac{3}{2\left( 1 - \alpha\right)}} .
\end{equation}
In particular, we see that a vanishing effective cosmological constant
enforces vanishing $\lambda$. In this limit the ``self-tuning''
mechanism breaks down.   
Thus, it seems that the most natural way to solve the instability is by 
allowing a positive 4d cosmological constant $L_4$. A consistent model 
thus predicts a positive value of $\Lambda_{\text{eff}}$ as a curved
solution in de Sitter space.
As we shall see next, there 
are indeed nearby curved solutions with such a positive 4d cosmological 
constant.

\section{Nearby Curved Solutions \label{sec:curved}}   

We consider now nearby curved solutions to the system in (\ref{eq:bulkaction}). 
From now on we can simply focus on the case of a real scalar field without loss of generality, since the final conclusions are independent of the reality of the field. 
The 5d metric now takes the form

\be
ds^2 = a(y)^2 g_{\mu\nu} \,dx^\mu dx^\nu  + dy^2
 \ee
where  $g_{\mu\nu}$ is maximally symmetric with 4d cosmological
constant given by $L_4$.  
The Einstein and scalar field equations of motion are modified in this case, becoming\footnote{Note that in the real case, we have also replaced $\lambda\to \lambda/2$.}: 
\bea
3\, \frac{a''}{a} + 3\,\left(\frac{a'}{a}\right)^2- 3 \, L_4  &=&  
		- \frac{\lambda\kappa_5^2}{2}\,(\varphi')^{2\alpha} - f\kappa_5^2\,\delta(y)  \label{E1}\\
6\left(\frac{a'}{a}\right)^2 - 6\,\frac{L_4}{a^2} &=& \frac{\lambda\kappa_5^2}{2}\,(2\alpha -1)\,  (\varphi')^{2\alpha} 
						\label{E2}\\
(2\alpha -1)\, \frac{\varphi''}{\varphi'} + 4\, \frac{a'}{a} &=& 
		\frac{1}{\alpha\lambda}\,\frac{df}{d\varphi}\,(\varphi')^{1-2\alpha}\,\delta(y) \label{dilaton}
\eea
The bulk equations can be solved explicitly in terms of Hypergeometric functions ${}_2F_1$ as follows. From (\ref{dilaton}) we find that 
\be\label{dilsol}
\varphi' = C_1 \, a^{4/(1-2\alpha)}
\ee
with $C_1>0$. Using this into (\ref{E2}) we can write
\be
a'^2 =    L_4\, +  \frac{\lambda \kappa_5^2}{12}\,(2\alpha-1)\,C_1^{2\alpha}  \, a^{2\Gamma} \, 
\ee
with $\Gamma = \frac{1+2\alpha}{1-2\alpha}$.
Integration of this gives as solution
\be
\int{\frac{d a}{\sqrt{L_4+\frac{\lambda\,\kappa_5^2\,(2\alpha-1)\,C_1^{2\alpha}}{12}\,a^{2\Gamma}}}} = \frac{a}{\sqrt{L_4}} \,{}_2F_1\left[\frac{1}{2\Gamma},\frac{1}{2},1+\frac{1}{2\Gamma},-\frac{\lambda\kappa_5^2\,(2\alpha-1)}{12}\frac{a^{2\Gamma}}{L_4}\right] \,.
\ee
Using this solution one can plug it into (\ref{dilsol}) to obtain a solution for the scalar field too. 
 One can check that real solutions exist for $L_4>0$, de Sitter, for both cases, $\lambda\,(2\alpha-1)$ positive or negative. Thus in principle curved solutions are not excluded in general\footnote{Solutions including a bulk cosmological constant can also be found following the same steps as above and are also given in terms of Hypergeometric functions.}.

In a completely consistent discussion of our stabilised solution,
$L_4$ should replace the finite value $y_c$. In particular, it should
be possible to take $y_c$ to infinity without inducing a divergent
instability as long as $L_4$ is non vanishing. Since it is very
complicated to study stability within the implicitly given nearby
curved solution we do not follow that route further here.

\section{Adding a Bulk Cosmological Constant}

We now discuss the possibility of adding a bulk cosmological constant to the action (\ref{eq:bulkaction}). So we consider now the system:
\be
S_{\text{tot}}=\int d^4x \, dy \,\sqrt{-G}\left(
 \frac{ R}{2\kappa_5^2} - \frac{\lambda}{2} \left( \partial_M \varphi \, \partial^M
    \varphi\right)^\alpha - \Lambda\ \right) -\int d^4 x\sqrt{-g}f\left(
  \varphi\right)_{\left| y=0\right.} .
\ee

Concentrating on flat solutions, the scalar field equation can be integrated just as before, so we get (\ref{dilsol}). Using this into the Einstein equations with a bulk cosmological constant we obtain

\be
a'^2 = \frac{\lambda  \kappa_5^2}{12}\,(2\alpha-1)\,C_1^{2\alpha}\, a^{2\Gamma} -
\frac{\Lambda\kappa_5^2}{6} \, a^2
\ee
where $\Gamma$ is defined as before. This equation can be integrated and the solutions for $a(y)$ read:
%
\be\label{asol2}
\hskip-0.04cm 
a^{\frac{4\alpha}{2\alpha-1}} \!= \!\left\{ \begin{array}{ll}
\!\!\!\sin{\left[\sqrt{\frac{\Lambda \kappa_5^2}{6}}\,\frac{4\alpha}{1-2\alpha}(| y| + C_2) \right]}  \sqrt{\frac{\lambda(2\alpha-1)\,C_1^{2\alpha}}{2\Lambda}}& \textrm{for $\lambda(2\alpha-1)>0$ and $\Lambda>0$} \\
\!\!\!\sinh{\left[\sqrt{\frac{-\Lambda \kappa_5^2}{6}}\,\frac{4\alpha}{1-2\alpha}(|y| + C_2) \right]} \sqrt{\frac{\lambda(2\alpha-1)\,C_1^{2\alpha}}{-2\Lambda}}&  \textrm{for $\lambda(2\alpha-1)>0$ and $\Lambda<0$} \\
\!\!\!\cosh{\left[\sqrt{-\frac{\Lambda \kappa_5^2}{6}}\,\frac{4\alpha}{1-2\alpha}(|y| + C_2) \right]} \sqrt{\frac{\lambda(2\alpha-1)\,C_1^{2\alpha}}{2\Lambda}} &  \textrm{for $\lambda(2\alpha-1)<0$ and $\Lambda<0$} \end{array}
\right.
\ee

\smallskip

Again, using this solution, one can find the solution for the scalar field using (\ref{dilsol}). 
Just as in the $\Lambda=0$ case, also here we have various solutions depending on the value of $\alpha$ and the sign of $\lambda$. 
In order to determine which solutions can give a workable self-tuning mechanism, we can again compute the effective Planck mass using (\ref{Mplanck4}). 
Following the discussion in section 3, we have that the resulting Planck mass is proportional to the integral of
\be
M_p \propto \int_{-\infty}^{\infty}{dy \,a^2}
\ee
with $a(y)$ given in (\ref{asol2}) above. Therefore, in order to get a 
finite result without introducing singularities, and thus fine-tuning, 
we require $\alpha < 1/2$ 
(cf.~(\ref{eq:arange}) in the $\Lambda=0$ case) 
for the last two solutions in (\ref{asol2}) 
(the first does not yield a finite effective Planck mass for 
any value of $\alpha$).
 Hence, the ``self-tuning" mechanism could work for the  sinh solution
 with $\lambda <0$ and/or  the cosh solution with $\lambda >0$. The
 other solutions require the introduction of singularities in order to
 obtain a finite Planck mass. As has been discussed in the past
 \cite{Forste:2000ps,Forste:2000ft} this reintroduces  a fine-tuning.   
 We illustrate these results in Fig.~1.

\begin{figure}[h] 
\begin{center}
\includegraphics[scale=0.40]{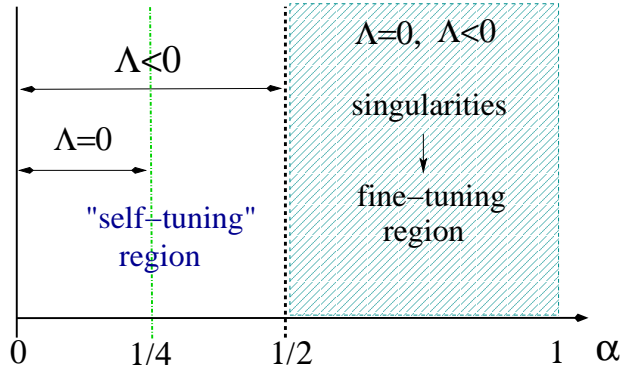}
\end{center}
\caption{In this figure we illustrate the regions of $\alpha$ where,
  barring stability issues,
  ``self-tuning'' mechanisms can be realised without introduction of
  singularities. The case with positive bulk
  cosmological constant is excluded for all values of $\alpha$.} 
\end{figure}

  We now follow our discussion in sections 4 and 5 in order to see if 
there is also in this case a problem with instabilities and, if so, 
estimate the amount of fine-tuning required. 

Taking into
account the leading exponential terms for large $y$ using the second
or third   
solution above (\ref{asol2}), gives an effective cosmological constant
of order  
\be
\Lambda_{\text{eff}} \sim M_p^4\, {\rm exp}
\left\{ -4 \,y_c \, \sqrt{-\frac{\Lambda\kappa_5^2}{6}} \right\}
\ee
which leads to the requirement 
\be\label{bound2}
{\rm exp}\left\{ 4\,y_c \sqrt{-\frac{\Lambda\kappa_5^2}{6}} \right\}> 
10^{120}
\ee
For the case $\lambda < 0$ (second solution in (\ref{asol2})), the  
stability analysis in section 5 can now be followed, again using
the leading contributions at large $y$ for the solution. The
requirement to cancel the divergent term coming from (\ref{flucs})
gives 
\be
\tilde \lambda^{1/\alpha} \,{\rm exp} \left\{-\frac{2(3-2\alpha)}{2\alpha-1}  \, y_c \sqrt{-\frac{\Lambda\kappa_5^2}{6}} \right\} <1
\ee
which implies, using (\ref{bound2}),  
\be
\tilde \lambda ^{1/\alpha} <  10^{-\frac{60(3-2\alpha)}{(1-2\alpha)}} ,
\ee
or expressed in terms of an effective cosmological constant
\begin{equation}
\tilde{\lambda}^{\frac{1}{\alpha}} <
\left(\frac{\Lambda_{\text{eff}}}{M_p^4}\right)^{\frac{3-
    2\alpha}{2(1-2\alpha)}} .
\end{equation}
Thus we see that also in this case, strong fine-tuning is needed in
order to obtain a stable result.  

For the case $\lambda > 0$ (third solution in (\ref{asol2})) the
argument is modified. Now, the $y$ dependent fluctuations come with
the wrong sign
\begin{equation}
S \supset \int d^4x \, dy \,\frac{\lambda \alpha \left( 1-2\alpha\right)
}{2} \, h( y) \left(\frac{d\delta \varphi}{dy}\right)^2 ,
\end{equation}
with 
\begin{equation}
h( y) = \left(\cosh{\left[\sqrt{-\frac{\Lambda
        \kappa_5^2}{6}}\,\frac{4\alpha}{1-2\alpha}(|y| + C_2) \right]}
\sqrt{\frac{\lambda(2\alpha-1)\,C_1^{2\alpha}}{2\Lambda}}\right)^{\frac{1}{\alpha}}  .
\end{equation}
This term gives rise to tachyonic KK masses as we see below. There can
be positive mass-squared terms from the expansion of the brane
Lagrangian. Without fine-tuning they should be of the order of the
Planck mass. To analyse the KK spectrum it is useful to redefine the
fifth coordinate such that
\begin{equation}\label{eq:ode}
\frac{du}{dy} = \frac{1}{h( y)} . 
\end{equation}
With this new coordinate we get a free Lagrangian
\begin{equation}\label{eq:KKval}
\int_{-y_c}^{y_c} dy\, h( y) \left(\frac{d\delta
    \varphi}{dy}\right)^2 = 
\int_{-u_c}^{u_c} du \left(\frac{d\delta \varphi}{du}\right)^2 .
\end{equation}
We periodically continue our set-up beyond the cut-off $u_c$. A KK
mode (with maximal amplitude on the brane) is,
\begin{equation}
\delta \varphi_n =a_n \cos \left( M_n u\right) ,
\end{equation}
with KK mass
\begin{equation}
M_n = \frac{2\pi n}{u_c} ,\,\,\, n \in {\mathbb Z} .
\end{equation}
If the KK masses were  not tachyonic it would be reasonable to
truncate the KK tower by a UV cut-off. For tachyonic KK modes one can
think of a formally similar mechanism. Tachyonic modes signal an
instability. They will stabilise by condensation and result in a
different solution to the equations of motion. However, if the wave
length of the condensed mode is much shorter than the shortest length
characterising our solution one might not notice the difference. In $u$
coordinates, the shortest length appearing in the solution
is\footnote{In $y$  coordinates this corresponds to a `cut-off' at the
  Planck mass, $ M_p > \left| p_y\right|  > \left| p_u\right|/h\left(
    y_c\right)$.} 
\begin{equation}\label{eq:wl}
u\left( y = \infty\right) - u\left( y = y_c\right) \sim \frac{1}{M_P}
\exp \left[ -\frac{4\,y_c}{1-2\alpha} \sqrt{-\frac{\Lambda
      \kappa_5^2}{6}}\right] ,
\end{equation}
where we assumed that all input parameters (except $\lambda$) are of
order one in Planck units. Now, consider a mode with wavelength,
$u_c/n \sim M_p^{-1}/n$, given by (\ref{eq:wl}). Integrating over the extra
dimension yields 
\begin{equation}
\int_{-y_c}^{y_c} dy\, h( y) \left(\frac{d\delta
    \varphi_n}{dy}\right)^2 \sim 
M_p \exp \left[\frac{8\,y_c}{1-2\alpha} \sqrt{-\frac{\Lambda
      \kappa_5^2}{6}}\right]\, a_n^2 .
\end{equation}
This tachyonic contribution can be balanced by contributions from
expanding the brane Lagrangian. Without fine-tuning such a term will
be of order one in Planck units. So, stabilisation of the relevant KK
modes is possible if
\begin{equation}
\tilde{\lambda}^{\frac{1}{\alpha}} \exp \left[\frac{8\,y_c}{1-2\alpha}\sqrt{\frac{-\Lambda
      \kappa_5^2}{6}}\right] < 1 .
\end{equation}
With the estimate for the lower bound on $y_c$ (\ref{bound2}) we
obtain a fine-tuning relation
\begin{equation}
\tilde{\lambda}^{\frac{1}{\alpha}} < 10^{-\frac{240}{1-2\alpha}} .
\end{equation}
Again we can express this as a function of the effective cosmological
constant
\begin{equation}
\tilde{\lambda}^{\frac{1}{\alpha}} < \left(
  \frac{\Lambda_{\text{eff}}}{M_p^4}\right) ^{\frac{2}{1 - 2\alpha}} .
\end{equation}

\section{Relation to Models with Three-Form Potential}

In this section we relate our models to dual configurations where the scalar
is replaced by  a three-form gauge potential $A_{\it 3}$. In particular we
see that the model discussed in \cite{Kim:2001ez} is dual to the case
$\alpha = 1/3$, as already observed in \cite{Choi:2002pp}. Focusing on the three-form, we consider the following 5d Lagrangian
\begin{equation}
S_H = \int d^5 x \sqrt{-G}\, \frac{\lambda}{2\beta}\left(
  -\frac{H_{\it 4}^2}{4!}\right)^\beta ,
\end{equation}
with $H_{\it 4}$ being an exact four-form\footnote{In the Euclidean
  continuation one should replace $H_{\it 4} \to i H_{\textit{4}} $.} 
\begin{equation}
H_{NOPQ} =\partial_{\left[ N\right. } A_{\left. OPQ\right]} .
\end{equation}
Numerical factors are chosen for convenience. Instead
of considering variations with respect to the three-form $A_{\it 3}$ we can
enforce the Bianchi identity with a Lagrange multiplier term
\begin{equation}
S_{\text{LM}} = -\int d^5 x\, \varphi\, \epsilon^{MNOPQ}\,\partial_M
H_{NOPQ}
\end{equation}
and take variations of the sum, $S_H + S_{\text{LM}}$, with respect to
  the four-form, $H_{\it 4}$, and the Lagrange multiplier, $\varphi$. The
  resulting equations of motion are
\begin{align}
\partial_M \varphi\,\epsilon^{MNOPQ} & = \sqrt{-G}\lambda \left(
  -\frac{H_{\it 4}^2}{4!}\right)^{\beta -1} H^{NOPQ} ,\label{eq:Hphi}\\
\partial_{[M}H_{NOPQ]} & = 0.
\end{align}
The equation obtained by varying $S_H$ w.r.t.\ $A_{\it 3}$ is reproduced as the
Bianchi identity, $\partial_{[M}\partial_{N]} \varphi =0$ and  (\ref{eq:Hphi}). To
obtain the dual theory we solve the algebraic equation for $H_{\it 4}$
(\ref{eq:Hphi}) by
\begin{equation}
H^{NOPQ} = \frac{\lambda^{-\frac{1}{2\beta
      -1}}}{\sqrt{-G}}\left( \partial_K\varphi \partial^K
  \varphi\right)^{\frac{1-\beta}{2\beta -1}}\epsilon^{MNOPQ}\partial_M
\varphi .
\end{equation}
Plugging this into the action, $S_H + S_{\text{LM}}$ provides the dual
action
\begin{equation}
\tilde{S} = \int d^5x \sqrt{-G} \left(\frac{1 -
  2\beta}{2\beta}\right)\lambda^{-\frac{1}{2\beta
    -1}}\left( \partial_M\varphi \partial^M
  \varphi\right)^{\frac{\beta}{2\beta -1}} .
\end{equation}
In particular we see that the model in \cite{Kim:2001ez} is dual to
our model with $\alpha = 1/3$ and a bulk cosmological
constant. 
In fact, the solution in \cite{Kim:2001ez} corresponds precisely to our third solution in the previous section with $\alpha = 1/3$, negative bulk cosmological constant and $\lambda(2\alpha- 1) <0$. 
Therefore  the instability problem we have discussed above,  has to be
addressed in this self-tuning proposal as well, as already observed    
in \cite{Medved:2001ad}. With the results presented here this would then
lead to a system with a positive cosmological constant, subject to a 
fine-tuning to be consistent with observations.

\section{Conclusions}

The possibility that the cosmological constant can be set to  (almost) zero by means of  extra dimensions, is definitely a very attractive idea. However, so far, all known self-tuning mechanisms of the cosmological constant in extra dimensions have been shown to require hidden fine-tunings, once closer analysis  of the mechanism is performed. 
In the present paper we have discussed a self-tuning mechanism in five dimensions, which makes use of non standard kinetic terms for the self-tuning fields. 
 
Starting with a (complex or real) scalar field with an unorthodox Lagrangian in five dimensions, we have shown that apparently consistent self-tuning solutions arise naturally, which can compensate and cancel the (classical part of the)  4d cosmological constant. 
A closer look revealed, unfortunately, that fluctuations around the 
self-tuning solution destabilise the mechanism, unless severe 
fine-tuning is reintroduced to obtain a small cosmological constant. 
Addition of a bulk cosmological constant does not improve the situation 
and the instability, or fine-tuning, persists, as we have shown. 

Quite surprisingly, the conditions on our coupling  $\lambda$ look
very model dependent (on $\alpha$ and a bulk cosmological constant).
Here, however we should notice that these are estimates. An accurate
treatment should follow the discussion in section
\ref{sec:curved}. Further, the condition on the smallness is not yet a
fine-tuning condition. 
Fine-tuning means that bare quantities have to
be chosen very precisely to cancel quantum corrections. 
The amount of fine-tuning is thus related to the cut-off dependence of
quantum corrections. It might happen that taking into
account anomalous dimensions of $\lambda$ results finally in a
universal fine-tuning condition of the same amount as in the original
cosmological constant problem.

Using our analysis, we have also demonstrated that the apparent self-tuning 
mechanism proposed in \cite{Kim:2001ez} is a particular case of our general 
class of models. This can be easily understood from the 
fact that the model  in \cite{Kim:2001ez} uses a three-form potential, which 
is dual to a scalar field in five dimensions. 
Thus we  have seen that so far a model with a fully consistent
self-tuning mechanism (without a hidden fine-tuning) does not
exist yet in five dimensions. We expect that similar  
type of problems might
arise in higher dimensions.

\section*{Acknowledgements}
This work was partially supported by the SFB-Transregio TR33 ``The Dark
Universe" (Deutsche Forschungsgemeinschaft) and the European Union 7th
network program ``Unification in the LHC era" (PITN-GA-2009-237920).

\end{document}